\documentclass[prl,twocolumn,superscriptaddress,floatfix,showpacs]{revtex4}
\usepackage{amsmath,amsfonts,epsf,dcolumn,natbib,graphicx}
\usepackage{ascii}
\usepackage{xcolor}
\usepackage{soul}
\usepackage{ulem}
\newcommand{\half}{\tfrac{1}{2}}

\newcommand{\be}{\begin{equation}}
\newcommand{\ee}{\end{equation}}
\newcommand{\beq}{\begin{equation}}
\newcommand{\eeq}{\end{equation}}
\newcommand{\bea}{\begin{eqnarray}}
\newcommand{\eea}{\end{eqnarray}}

\begin{document}
\bibliographystyle{plainnat}
%
\title{
Local Spin-density Approximation  \\
Exchange-correlation Free-energy Functional 
}

\author{Valentin V.~Karasiev}
\email{vkarasev@qtp.ufl.edu}
\affiliation{Quantum Theory Project, 
Departments of Physics and of Chemistry,
University of Florida, Gainesville FL 32611-8435}
\author{Travis Sjostrom}
\affiliation{Theoretical Division, Los Alamos National Laboratory, Los Alamos, NM 87545}
\author{James Dufty}
\affiliation{Department of Physics, 
University of Florida, Gainesville FL 32611-8435}
\author{S.B.~Trickey}
\affiliation{Quantum Theory Project, 
Departments of Physics and of Chemistry,
University of Florida, Gainesville FL 32611-8435}

\date{November 18, 2013; revised January 13, 2014}
\begin{abstract}
\noindent An accurate analytical parametrization for 
the exchange-correlation free energy
of the homogeneous electron gas, including interpolation for
partial spin-polarization, is 
derived via thermodynamic analysis of recent
restricted path integral Monte-Carlo (RPIMC) data.  
This parametrization constitutes the local spin density approximation (LSDA) 
for the exchange-correlation 
functional in density functional theory. The new
finite-temperature LSDA reproduces the RPIMC data well, satisfies the 
correct high-density and low- and high-$T$ asymptotic limits, and is
well-behaved beyond the range of the RPIMC data, suggestive of broad
utility.
\end{abstract}


\maketitle

The homogeneous electron gas (HEG) is a fundamentally important
system for understanding many-fermion physics.  In the
absence of exact analytical solutions for its energetics, high-precision
numerical results have been critical to insight.  
Recently published \cite{Brown.PRL} restricted path integral Monte Carlo
(RPIMC) data for the HEG over a
wide range of temperatures and densities open the opportunity
to obtain closed form expressions for HEG thermodynamics, in 
particular the exchange and correlation (XC) contributions.    
Such expressions extracted from Monte Carlo data are well-known for the 
zero-$T$ HEG, where they have played a major role in understanding
inhomogeneous electron-system behavior. We provide the 
corresponding thermodynamical expressions for wide temperature and density
ranges.  

Density functional theory (DFT) is the motivating context.  
For ground-state DFT, 
the most basic exchange-correlation (XC) density functional 
is the local density approximation (LDA).  It approximates the 
local XC energy per particle, $\varepsilon_{\rm xc}$, as 
the value for the HEG at the local density, 
$\varepsilon_{\mathrm{xc}}^{\mathrm{LDA}}(n({\mathbf r})) \approx %
\varepsilon_{\mathrm{xc}}^{\mathrm{HEG}} (n)|_{n=n({\mathbf r})}$ [also see 
Eq.\ (\ref{FxcLSDA}) below]. 
Computational implementation is via 
parametrizations \cite{PZ81,VWN80} of 
HEG quantum Monte Carlo (QMC) data \cite{CeperleyAlder.1980}. Recent
QMC results \cite{SND.2013} for the spin-polarized $T=0$ K  HEG  
also validate the spin-interpolation formulae used
in that case, the local spin density approximation (LSDA). 
All more refined $\varepsilon_{\mathrm{xc}}$ approximations reduce
to the LSDA in the weak inhomogeneity limit.  

Finite-temperature DFT 
\cite{Mermin65,Stoitsov88,Dreizler89} 
increasingly is being used to study matter under diverse
density and temperature conditions 
\cite{Holstetal,Lambert06,Surh2001,PRE.86.056704.2012,VT84F,Hu.Militzer..2011}. 
In it, the XC free-energy is defined by  
decomposition of the universal free-energy density functional 
(independent of the external potential).  With the 
$T$-dependence suppressed for now, that functional is 
\be
{\mathcal F}[n]={\mathcal T}_{\mathrm{s}}[n]-T{\mathcal S}_{\mathrm{s}}[n]
+{\mathcal F}_{\mathrm H}[n]+{\mathcal F}_{\mathrm{xc}}[n]
\, .
\label{I1}
\ee
The first two terms are the non-interacting kinetic energy and 
entropy (also known as the Kohn-Sham KE and entropy), 
${\mathcal F}_{\mathrm H}[n]$  is 
the classical electron-electron Coulomb energy,
and the XC free energy by definition is
\bea
{\mathcal F}_{\mathrm {xc}}[n]& := & ({\mathcal T}[n]- %
{\mathcal T}_{\mathrm{s}}[n]) - T({\mathcal S}[n]- %
{\mathcal S}_{\mathrm{s}}[n]) \nonumber \\
&&+  ({\mathcal U}_{\mathrm {ee}}[n]-{\mathcal F}_{\mathrm H}[n])\,  ,
\label{I2}
\eea
with  ${\mathcal T}[n]$ and ${\mathcal S}[n]$ the interacting system 
kinetic energy and entropy and ${\mathcal U}_{\mathrm {ee}}[n]$ 
the full
quantum mechanical electron-electron interaction energy.

Just as for $T = 0$ K, the existence theorems of finite-$T$ DFT are 
not constructive for $\mathcal{F}_{\mathrm{xc}}$, so approximations 
must be devised. 
Common practice \cite{Holstetal} in simulations is to use 
a $T = 0$ K XC functional, 
${\mathcal F}_{\mathrm{xc}}[n(T),T] \approx E_{\mathrm{xc}}%
[n(T)]$.  This   
gives only the implicit $T$-dependence provided by $n({\mathbf r},T)$. 
However, there is substantial evidence from both finite-$T$ Hartree-Fock 
 \cite{PRE.86.056704.2012,PRB85-045125-2012} and 
finite-$T$ exact exchange calculations 
\cite{LippertModineWright2006,GreinerCarrierGoerling2010}
of non-negligible $T$-dependence in exchange itself.

Addressing that $T$-dependence until now has been hampered by lack of
an accurate, simulation-based LDA for ${\mathcal F}_{\mathrm{xc}}$.  Thus, 
several 
${\mathcal F}_{\mathrm{xc}}$ approximations have been proposed 
on the basis of various models; see Ref.\ \onlinecite{SD.2013} 
and references therein.  The RPIMC data for the HEG in
Ref.\ \onlinecite{Brown.PRL} provide the opportunity to fill that gap
with an LSDA on equivalent footing with the ground state 
$E_{\mathrm {xc}}^{\mathrm{LSDA}}$.
Note that Ref.\ \onlinecite{Brown.fit} provided a fit 
for the RPIMC XC {\it internal} energy data but not for 
${\mathcal F}_{\mathrm{xc}}$.  Subsequently, an error in that fit was 
corrected.  Here we use the corrected fit, denoted ``BDHC''.  
Incidentally to the main 
theme of Ref.\ \onlinecite{SD.2013}, two of us fitted the 
unpolarized finite-$T$ RPIMC results \cite{Brown.PRL} 
and extracted a parametrization of the HEG XC free energy. 
That constitutes an LDA ${\mathcal F}_{\mathrm{xc}}$.
But several important issues were not treated,  namely which 
of several possible thermodynamic routes is optimal for 
extracting ${\mathcal F}_{\mathrm {xc}}$, what functional form is 
most reliable for the requisite fitting of the RPIMC data,
which RPIMC data to use, and how to handle the partially polarized 
case.  We address those here to provide 
the free energy LDA and LSDA with full $T$-dependence,
\begin{align}
\mathcal{F}_{\mathrm{xc}}[n(T),T]  & =\int d\mathbf{r}n\left( \mathbf{r},T\right)
f_{\mathrm{xc}}[n(T);\mathbf{r}, T]\nonumber \\
 \approx\int d\mathbf{r}n \left(  \mathbf{r},T\right)&  f_{\mathrm{xc}}%
^{\mathrm{HEG}}(n\left(  \mathbf{r},T\right)  ,T)\equiv\mathcal{F}_{\mathrm{xc}}^{\mathrm{LSDA}}%
[n(T),T]
\label{FxcLSDA}
\end{align}
where $f_{\mathrm{xc}}^{\mathrm{HEG}}(n,T)=F_{\mathrm{xc}}^{\mathrm{HEG}}(n,T)/N$ 
is the 
XC free energy per particle for the HEG and $N$ the electron number. 
Note that at
$T=0$ K, $f_{\mathrm{xc}}=\varepsilon_{\mathrm{xc}}$ and ${\mathcal F}_{\mathrm{xc}}= E_{\mathrm{xc}}$.

Unless noted otherwise, we use Hartree atomic units. (Observe that 
Refs.\ \onlinecite{Brown.PRL} and \onlinecite{Brown.fit} use Rydberg au.)
The interacting HEG is described completely by three parameters, the density 
$n^{\mathrm{HEG}}=n=N/V$,
spin-polarization $\zeta =  (n_{\mathrm {\uparrow}}- n_{\mathrm{\downarrow}})/n$,  
 and temperature $T$.   Its XC 
free energy per particle,
$f_{\mathrm{xc}}^{\mathrm{HEG}}(n^{\mathrm{HEG}},T) = {\mathcal F}_{\mathrm{xc}}^{\mathrm{HEG}}[n^{\mathrm{HEG}},T]/N$, 
is the quantity of interest. 
As usual, we use the Wigner-Seitz radius, $r_s=(3/4\pi n)^{1/3}$, and reduced 
temperature $t=T/T_F$, with the Fermi temperature
$T_F^{\zeta=0} =[3\pi^2n]^{2/3}/2k_B$ for the unpolarized case and 
$T_F^{\zeta=1} =[6\pi^2n]^{2/3}/2k_B$ for the fully polarized case. Significant densities
range from $r_s\ll 1$ through $r_s \ge 10$.  The relevant temperature
range is at least $0 \le t \le 10$. While large $t$ represents the 
classical limit, 
the approach to it will vary with $r_s$, via the dimensionless Coulomb 
coupling parameter, $\Gamma=2\lambda^2 r_{\mathrm{s}}/t$ with $\lambda=(4/9\pi)^{1/3}$. 

The  RPIMC data for the HEG \cite{Brown.PRL} 
are the total kinetic ${\mathcal T}$ and potential (or interaction) 
${\mathcal U}_{\mathrm{ee}}$ energies for given $r_s$ and $t$. 
The issues are which RPIMC data to use and how best to 
extract a broadly reliable $f_{\mathrm{xc}}$ from that data.

One thermodynamic route is  
via the RPIMC data for the XC internal energy per particle, which is the 
difference of the 
interacting and non-interacting system total internal energies per particle, 
$\varepsilon_{\rm xc}=\tau+u_{\mathrm{ee}}-\tau_{\mathrm{s}}$, 
with $\tau={\mathcal T}/N$, $u_{\mathrm{ee}}={\mathcal U}_{\mathrm{ee}}/N$, 
and  
$\tau_{\mathrm{s}}={\mathcal T}_{\mathrm{s}}/N$ the non-interacting HEG 
kinetic energy per particle  
({\it i.e.}, ${\mathcal T}_{\mathrm s}$ is the finite-$T$ Thomas-Fermi 
KE \cite{Feynman..Teller.1949,KST2}).  Observe that $\tau_{\mathrm{s}}$ is 
given both analytically and tabularly in the Supplementary Material for 
Ref.\ \onlinecite{Brown.PRL}. 
From Eq.\ (\ref{I2}) $f_{\rm xc}=\varepsilon_{\rm xc}-T\sigma_{\rm xc}$ 
which, with a standard thermodynamic relation for the entropic contribution
per particle
\be
\sigma_{\rm xc}(r_{\mathrm{s}},t)=-\frac{t}{T}\frac{\partial f_{\rm xc}(r_{\mathrm{s}},t)}%
{\partial t}\Big|_{r_{\mathrm{s}}}\,,
\label{E1}
\ee
gives
\be
f_{\rm xc}(r_{\mathrm{s}},t)-t\frac{\partial f_{\rm xc}(r_{\mathrm{s}},t)}{\partial t}\Big|_{r_{\mathrm{s}}}
=\varepsilon_{\rm xc}(r_{\mathrm{s}},t)\,.
\label{E2}
\ee
Observe that the ${\mathcal F}_{\mathrm H}$ from (\ref{I2}) vanishes for
the HEG because of the neutralizing background.  

Reference  \onlinecite{SD.2013} used another thermodynamic relation to 
obtain $f_{\rm xc}$ directly from the RPIMC interaction energy $u_{\mathrm{ee}}$
per particle via integration over  $\Gamma$, the coupling constant 
\cite{TMI.1985}.  This is equivalent \cite{Schweng.Boehm.1993} to 
\be
f_{\mathrm{xc}}(r_{\mathrm{s}},t)=\frac{1}{r_{\mathrm{s}}^2}
\int_0^{r_{\mathrm{s}}} dr^{\prime}_{\mathrm{s}} r^{\prime}_{\mathrm{s}} u_{\mathrm{ee}}(r^{\prime}_{\mathrm{s}},t)|_t
\,.
\label{fxc-int}
\ee
Exact integration requires the choice of an integrable form fitted to 
the RPIMC data for $u_{\mathrm{ee}}$.  Instead, differentiation of 
Eq.\ (\ref{fxc-int}) with respect to $r_{\mathrm{s}}$ gives \vspace*{-6pt}
\be
2f_{\rm xc}(r_{\mathrm{s}},t)+
r_{\mathrm{s}}\frac{\partial f_{\rm xc}(r_{\mathrm{s}},t)}{\partial r_{\mathrm{s}}}\Big|_t
=u_{\mathrm{ee}}(r_{\mathrm{s}},t)\;,
\label{fxc-eint}
\vspace*{-6pt}
\ee
which is the analogue of Eq.\ (\ref{E2}). 
Eqs.\ (\ref{E2}) and (\ref{fxc-eint}) may be combined to yield \vspace*{-6pt}
\begin{align}
\tau_{\mathrm{s}}(r_{\mathrm{s}},t) 
&-t\frac{\partial f_{\rm xc}(r_{\mathrm{s}},t)}{\partial t}\Big|_{r_{\mathrm{s}}}\nonumber \\
&-f_{\rm xc}(r_{\mathrm{s}},t)-r_{\mathrm{s}}\frac{\partial f_{\rm xc}(r_{\mathrm{s}},t)}{\partial r_{\mathrm{s}}}\Big|_t=\tau(r_{\mathrm{s}},t) \,.
\label{tau-fit}
\vspace*{-6pt}
\end{align}

Fitting a suitable analytical $f_{\mathrm {xc}}(r_{\mathrm{s}},t)$ to one 
of Eqs.\ (\ref{E2}), 
(\ref{fxc-eint}),  or (\ref{tau-fit}) constitutes our Fits A, B, and D 
respectively.  While Fits B and D each use only one subset of the 
RPIMC data ($u_{\mathrm{ee}}$, $\tau$ respectively), Fit A uses both 
via the combination 
$\varepsilon_{xc}$. A second way to use both data sets is to fit
 $f_{xc}$ to Eqs.\ (\ref{fxc-eint}) and (\ref{tau-fit}) concurrently; 
this is our Fit C.
All four Fits use the RPIMC data on its discrete mesh, while  
the assumed functional form for $f_{\mathrm{xc}}$ should provide
useful extrapolation outside the RPIMC data domain.   
Brown {\it et al.}\ used \cite{Brown.fit} a functional form similar to the 
Perrot-Dharma-wardana \cite{PDW2000} XC functional 
to fit the  RPIMC data for $\varepsilon_{\rm xc}$. We tested both the original
and Brown {\it et al.}\ 
versions and found physically 
implausible behavior (oscillations) in the $r_s$ dependence. 
See Supplemental Material \cite{SM}. 

A  Pad\'e approximant as originally given by 
Ichimaru \textit{et al.}\ \cite{TMI.1985,TI1986,TI.1989-I,Ichimaru.1993} 
and also employed in Ref.\ \onlinecite{SD.2013} for $u_{\rm ee}$
is suggestive. We used an extension of that form, but  
for $f_{\mathrm{xc}}$, for both the unpolarized and fully polarized 
cases. With explicit polarization labeling the form is\vspace*{-6pt}
\be
f_{\mathrm{xc}}^{\zeta}(r_{\mathrm{s}},t)=-\frac{1}{r_{\mathrm{s}}}
\frac{\omega_{\zeta} a(t)+b_\zeta(t)r_{\mathrm{s}}^{1/2}+c_\zeta(t)r_{\mathrm{s}}}
{1+d_\zeta(t)r_{\mathrm{s}}^{1/2}+e_\zeta(t)r_{\mathrm{s}}}
\,,
\label{fit2}
\vspace*{-6pt}
\ee
where $\omega_{0}=1$ and $\omega_{1}=2^{1/3}$ for $\zeta = 0$, $1$, respectively. 
The functions $a(t)$, $b_\zeta(t) - e_\zeta(t)$, in turn, are Pad\'e 
approximants in $t$.  
The original forms \cite{Ichimaru.1993} 
proved to be inadequate to reproduce the  
RPIMC $\varepsilon_{\mathrm{xc}}$ data at $r_{\mathrm{s}}=1$. This inflexibility
was remedied by adding one parameter in  $c_\zeta(t)$, with the resulting
definitions for $a(t)$, $b_\zeta(t) - e_\zeta(t)$  as follows
($\zeta$ labeling suppressed for clarity): 
\begin{align}
  a(t)=&0.610887 \tanh{ \left( \frac{1}{t} \right)  } \times \nonumber \\
  &\frac{0.75 + 3.04363 t^2-0.09227 t^3+1.7035 t^4}
  {1 + 8.31051 t^2 + 5.1105 t^4}
\label{a}
\\
  b(t)=&\tanh{ \left( \frac{1}{\sqrt{t}} \right)  }
  \frac{b_1 + b_2 t^2 + b_3 t^4}{1+b_4 t^2 +b_5 t^4} \\
  c(t)=&\left[ c_1 +c_2\exp \left( -\frac{c_3}{t} \right) \right] e(t) \\
  d(t)=&\tanh{ \left( \frac{1}{\sqrt{t}} \right)  }
  \frac{d_1 + d_2 t^2 + d_3 t^4}{1+d_4 t^2 +d_5 t^4} \\
  e(t)=& \tanh{ \left( \frac{1}{t} \right) }
  \frac{e_1 + e_2 t^2 + e_3 t^4}{1+e_4 t^2 +e_5 t^4}  \; .
\label{e}
\end{align}
In the small-$r_{\mathrm{s}}$ and small-$\Gamma$ limits, Eq.\ (\ref{fit2})  
reduces to the finite-$T$ X functional of Ref.\ \onlinecite{PDW84} 
(also see Refs.\ \onlinecite{TMI.1985}, \onlinecite{TI1986} for details),
\be
f_{\mathrm{x}}^{\zeta}(r_{\mathrm{s}},t)=-\frac{\omega_{\zeta}}{r_{\mathrm{s}}}a(t)
\,.
\label{fx}
\ee

\begin{table}
\caption{\label{tab:table1}
Fit A parameters for the XC free-energy functional for the unpolarized
($\zeta=0$) and fully polarized ($\zeta=1$) HEG.   
}
\begin{ruledtabular}
\begin{tabular}{lrr}
    & $\zeta=0$ & $\zeta=1$ \\
\hline
$b_{ 1}$ &  	0.283997     &      0.329001   \\
$b_{ 2}$ &     48.932154     &    111.598308   \\
$b_{ 3}$ &  	0.370919     &      0.537053   \\
$b_{ 4}$ &     61.095357     &    105.086663   \\
$b_{ 5 }$ & $\sqrt{3/2}~\lambda^{-1}b_3=$ 0.871837  &  
             $\sqrt{3/2}~2^{1/3}\lambda^{-1}b_3=$ 1.590438   \\
\hline
$c_{ 1}$ &  	0.870089     &      0.848930   \\
$c_{ 2}$ &  	0.193077     &      0.167952   \\
$c_{ 3}$ &  	2.414644     &      0.088820   \\
\hline
$d_{ 1}$ &  	0.579824     &      0.551330   \\
$d_{ 2}$ &     94.537454     &    180.213159   \\
$d_{ 3}$ &     97.839603     &    134.486231   \\
$d_{ 4}$ &     59.939999     &    103.861695   \\
$d_{ 5}$ &     24.388037     &     17.750710   \\
\hline
$e_{ 1}$ &  	0.212036     &      0.153124   \\
$e_{ 2}$ &     16.731249     &     19.543945   \\
$e_{ 3}$ &     28.485792     &     43.400337   \\
$e_{ 4}$ &     34.028876     &    120.255145   \\
$e_{ 5}$ &     17.235515     &     15.662836   \\
\end{tabular}
\end{ruledtabular}
\end{table}

The correct $T=0$ K
limit is obtained by using the recent $T=0$ K QMC data \cite{SND.2013}. Thus,
Eq.\ (\ref{fit2}) first was fitted at $t=0$ to the zero-$T$ QMC data.
That fixed the parameters $b_1$, $c_1$, $d_1$ and $e_1$.  The
remaining parameters in Eq.\ (\ref{fit2}) were fitted to the finite-$T$ 
RPIMC data.  The correct high-$T$ limit,
\be
\lim_{T\rightarrow \infty} f_{\mathrm{xc}}^{\zeta}(r_{\mathrm{s}},t)=-\frac{1}{\sqrt{3}}r_{\mathrm{s}}^{-3/2}T^{-1/2} 
+ O(T^{-1}) \,,
\label{highTlim}
\ee
for all $\zeta$,  
corresponds to the leading correlation term; 
see Refs.\ \onlinecite{DeWitt.1965,Perrot.1979,PDW84}.  It was 
incorporated by fixing the ratio between the parameters $b_{\zeta,5}=\sqrt{3/2}\omega_{\zeta}\lambda^{-1}b_{\zeta,3}$ 
in $b_{\zeta}(t)$.

Each of the thermodynamic routes, A, B, C, or D, to 
$f_{\mathrm{xc}}$ from 
an RPIMC data subset can be tested by computing values for both
the subset used in that fit and the unused subsets and comparing 
the results with the original RPIMC data.  For example, Fit A uses RPIMC  
$\varepsilon_{\mathrm{xc}}$ data as input to Eq.\ (\ref{E2}).  Thus,
we calculated 
values of $u_{\mathrm{ee}}^{\mathrm{fit}}$ via Eq.\ (\ref{fxc-eint})  
and $\tau^{\mathrm{fit}}$ via Eq.\ (\ref{tau-fit}) from the Fit A $f_{\mathrm xc}$ and compared the results 
to the RPIMC data in the form of mean absolute relative errors (MARE).  
The essential result is
that Fits A and C are close in quality but Fit A is modestly better
on grounds of MARE for $\varepsilon_{\mathrm{xc}}$.  From the
same perspective, the resulting fit to $\varepsilon_{\mathrm{xc}}$ also 
is better than the BDHC fit.  The final parameters are shown in Table \ref{tab:table1} and error 
comparisons are in Table \ref{tab:table2a}.  (Those parameters were done
with analytical derivatives in Eq.\ (\ref{E2}), after exploration of fits
with numerical thermodynamic derivatives.)
Other error comparisons are in the Supplemental Material \cite{SM}. \\[-7pt]

\begin{table}
\caption{\label{tab:table2a}
RPIMC data sets used for fits with MARE and absolute maximum 
relative errors (\%) 
for calculated kinetic, interaction, and XC {\it internal} energies per
particle for unpolarized  ($\zeta=0$) 
and fully polarized ($\zeta=1$) cases.
}
\begin{ruledtabular}
\begin{tabular}{lcccc}
Funct. &  fitted to &  $\tau$ & $u_{\mathrm{ee}}$ & $\varepsilon_{\mathrm{xc}}$ \\
\hline
\vspace{3pt}
\underline{$\zeta=0$} &&&&\\
BDHC &   $\varepsilon_{\mathrm{xc}}$ & - &- &  1.3/14 \\
Fit A &  $\varepsilon_{\mathrm{xc}}$ & 1.3/10  & 1.4/4.5 & 0.5/3.3 \\
Fit B &  $u_{\mathrm{ee}}$           & 1.8/6.1 & 0.3/1.2 & 1.9/9.2 \\          
Fit C &  $\tau~\&~ u_{\mathrm{ee}}$      & 1.0/8.3 & 0.5/2.8 & 1.2/7.5 \\
Fit D &  $\tau$                    & 0.6/5.1 & 5.0/18 & 5.6/23 \\
\hline
\vspace{3pt}
\underline{$\zeta=1$} &&&&\\
BDHC &   $\varepsilon_{\mathrm{xc}}$ & - &- &  2.3/18 \\
Fit A &  $\varepsilon_{\mathrm{xc}}$ & 1.7/13  & 1.6/4.8  & 1.2/7.8 \\
Fit B &  $u_{\mathrm{ee}}$           & 2.2/15 & 0.5/3.7 & 2.2/10 \\        
Fit C &  $\tau~\&~ u_{\mathrm{ee}}$  & 1.2/8.0 & 0.8/3.8 & 1.7/9.3 \\
Fit D  &  $\tau$                     & 0.6/4.2  & 6.3/17  & 7.2/25 \\
\end{tabular}
\end{ruledtabular}
\end{table}

\begin{figure*}
  \includegraphics*[width=5.5cm]{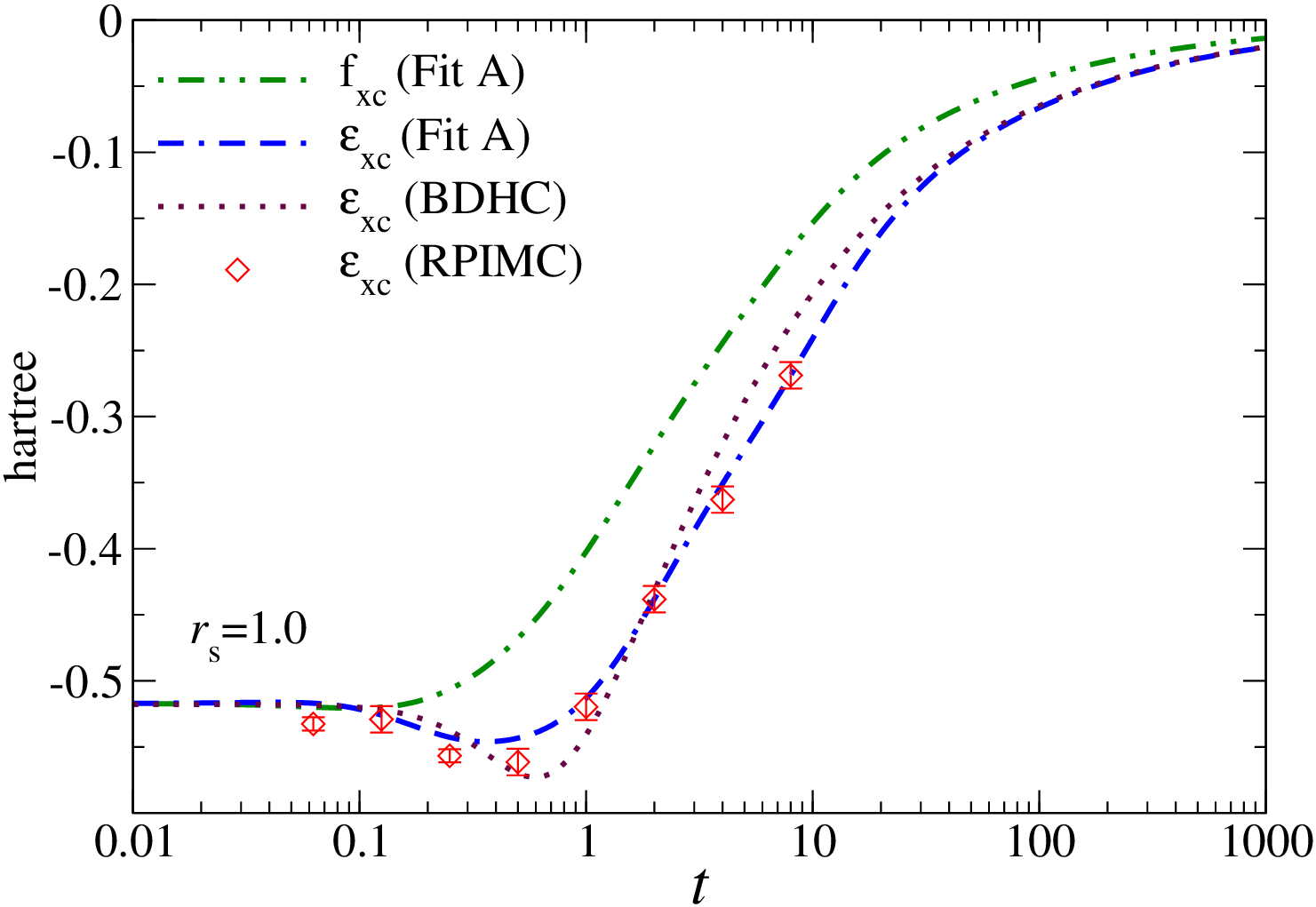}
  \includegraphics*[width=5.5cm]{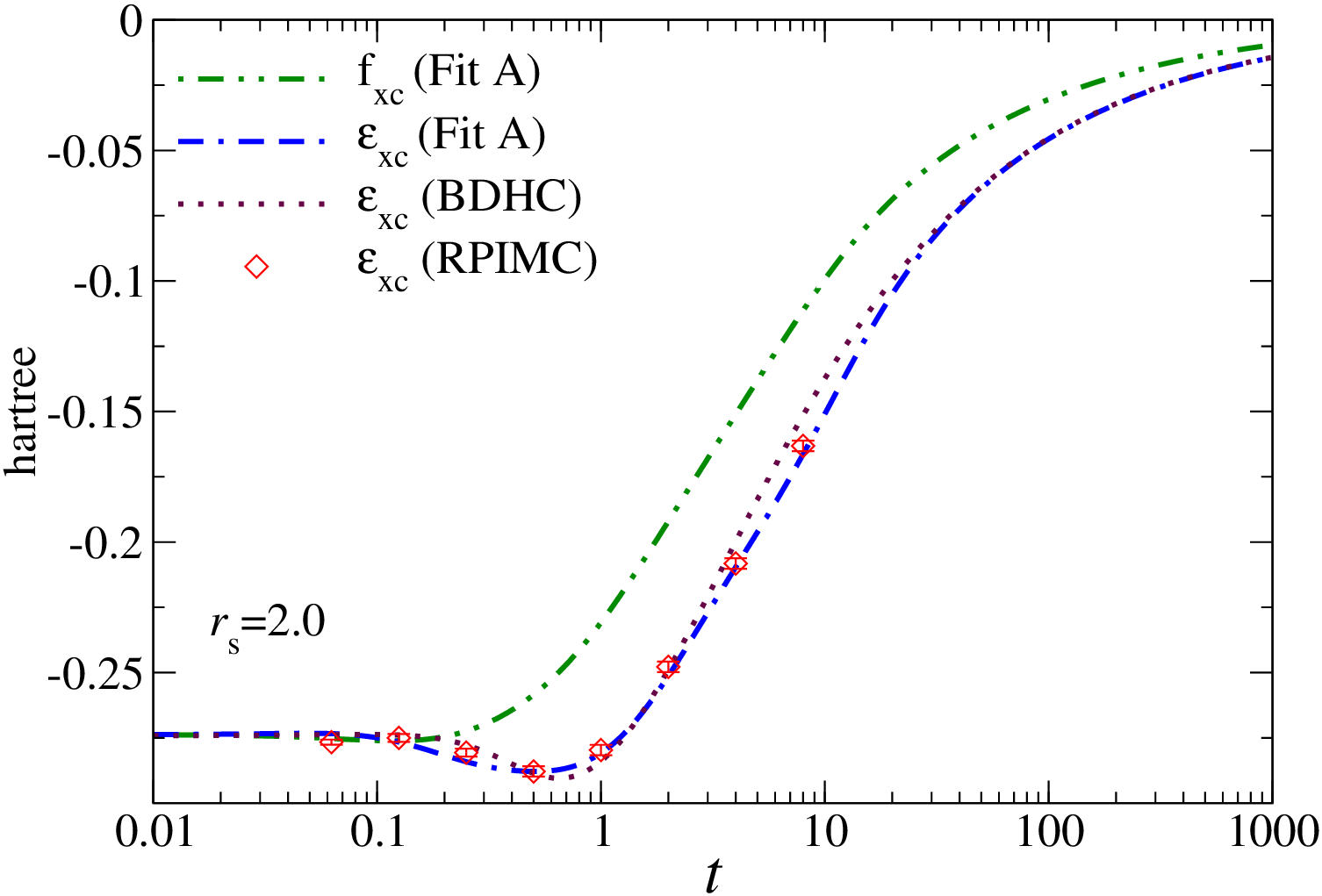}
  \includegraphics*[width=5.5cm]{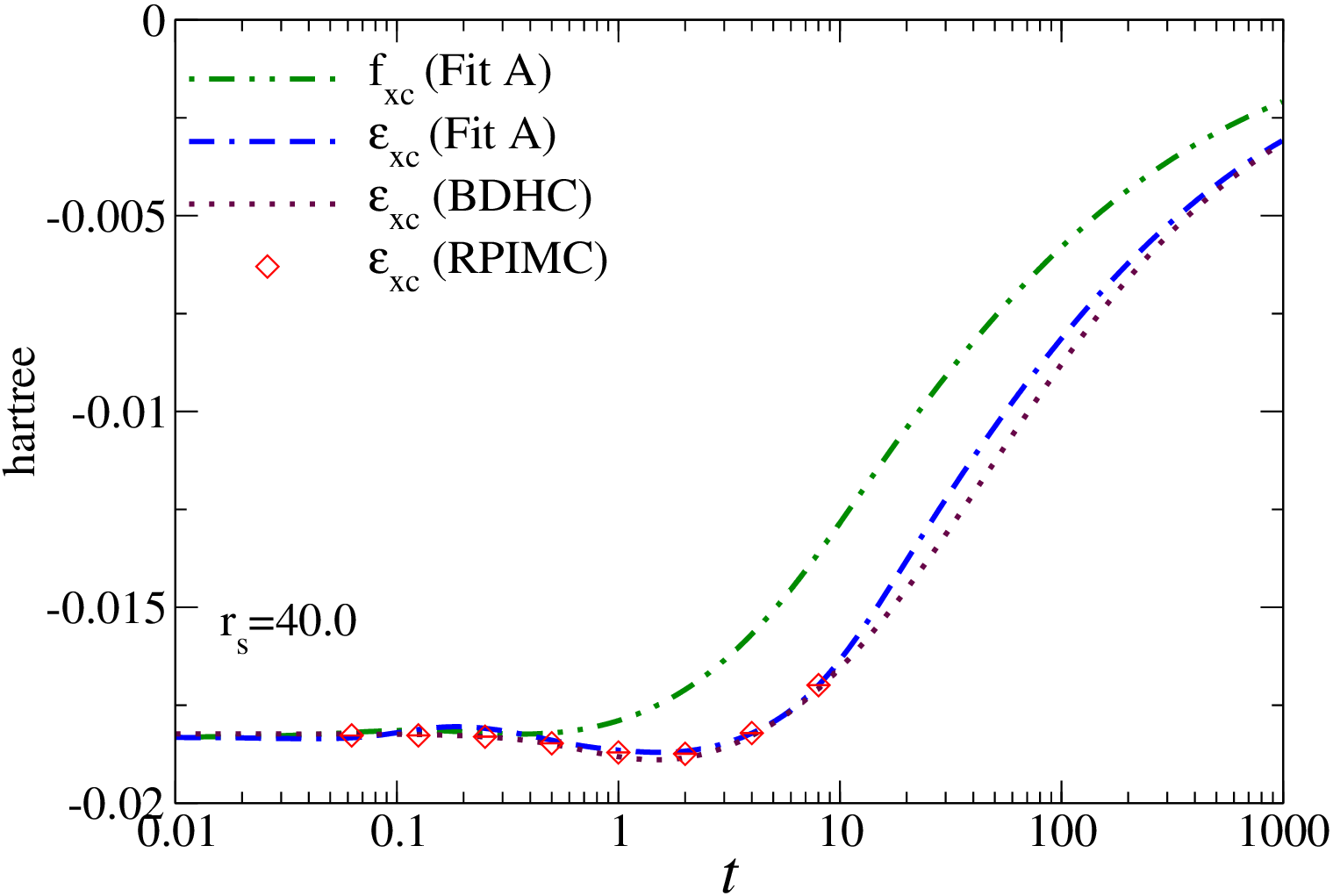}
  \includegraphics*[width=5.5cm]{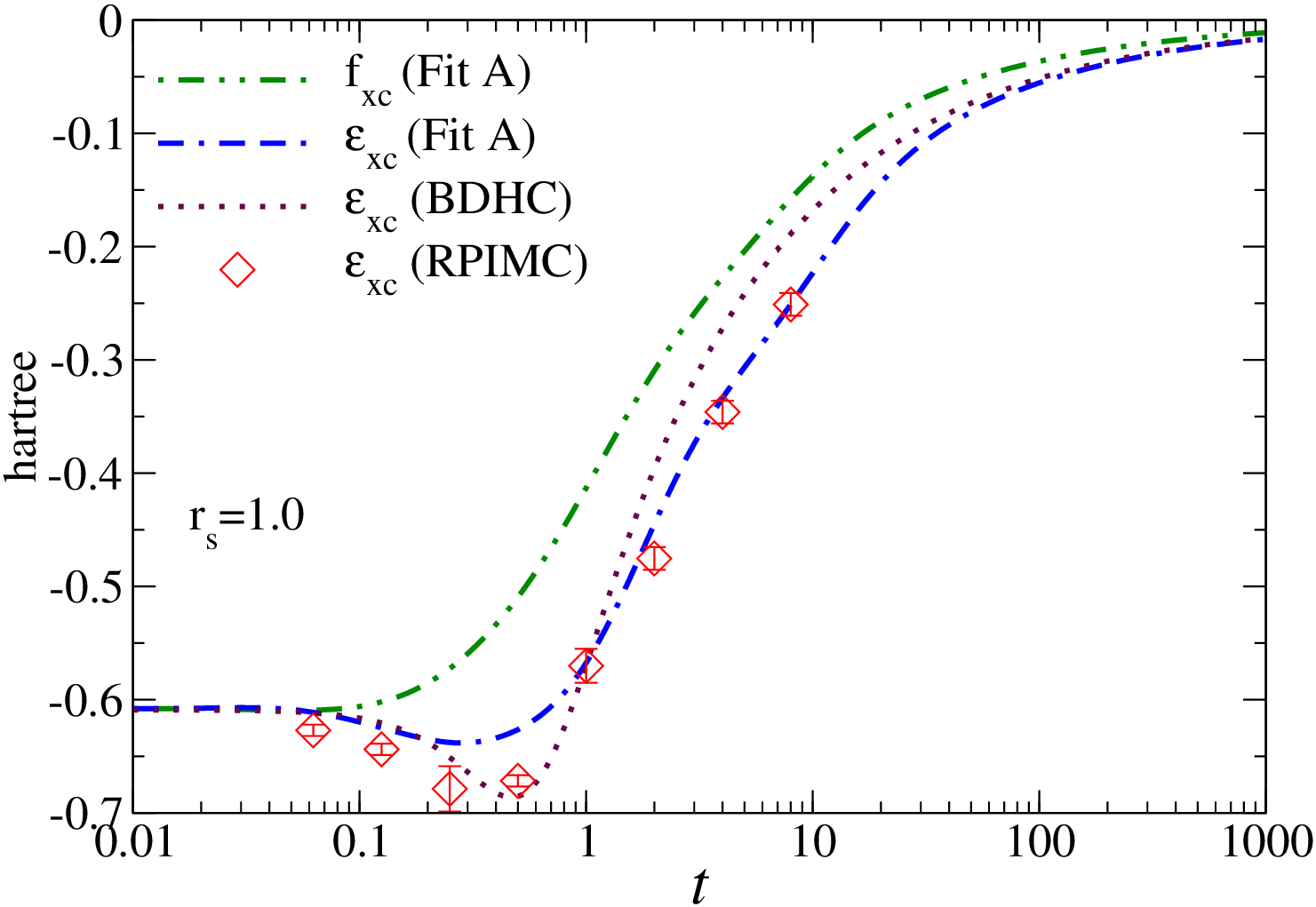}
  \includegraphics*[width=5.5cm]{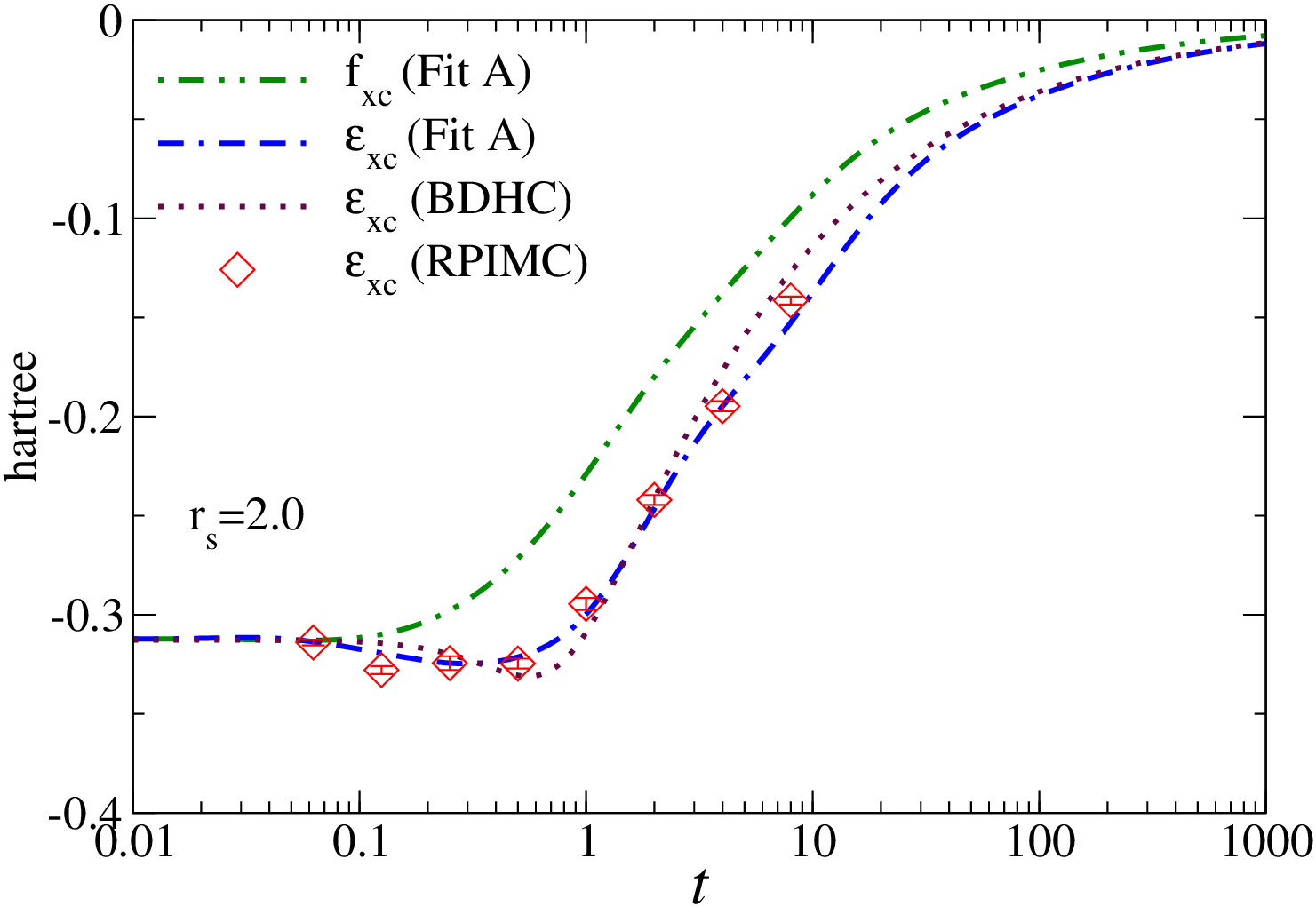}
  \includegraphics*[width=5.5cm]{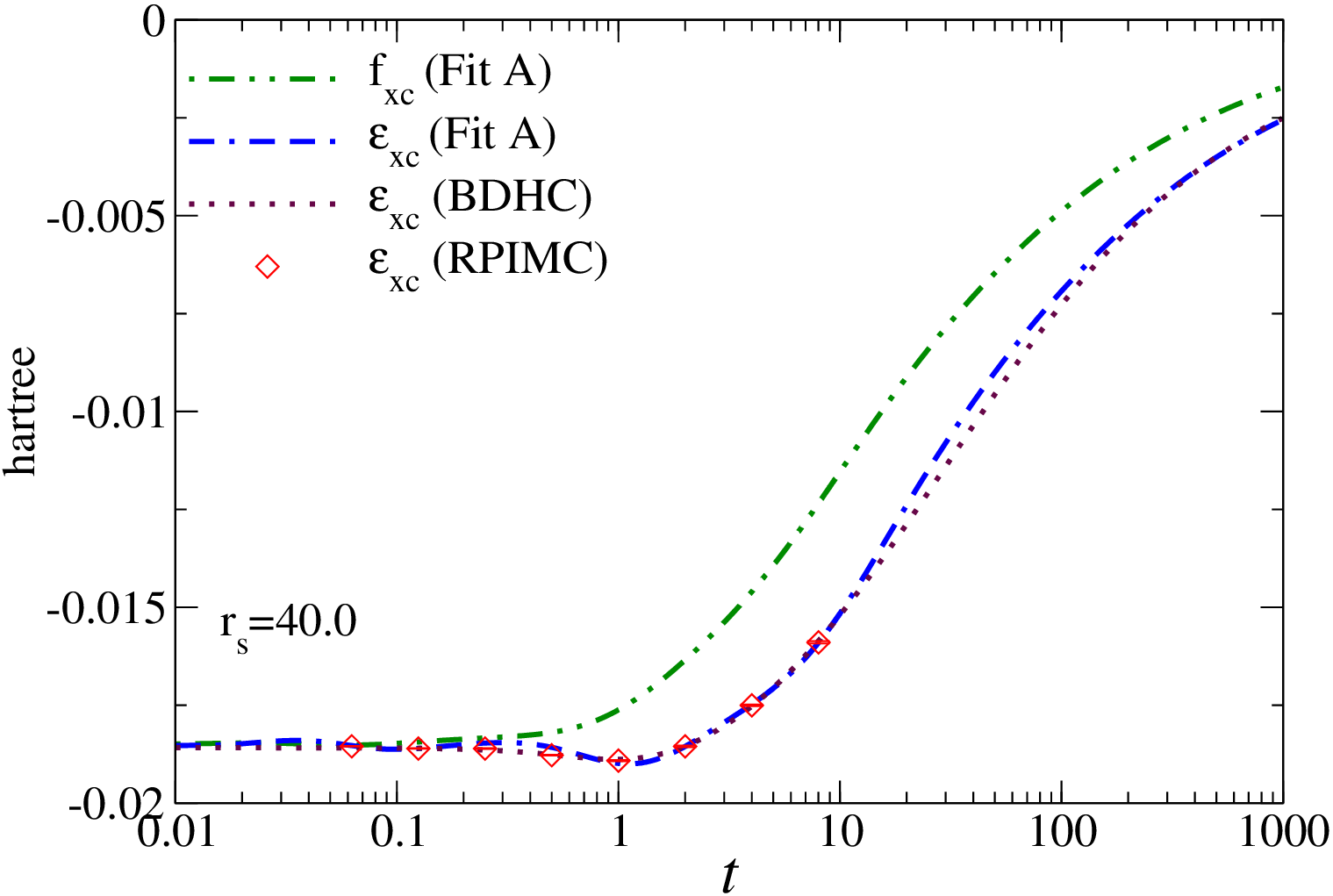}
 \caption{
 $\varepsilon_{\mathrm{xc}}$ and $f_{\mathrm{xc}}$  
 from Fit A for the unpolarized, $\zeta=0$ (top) and the fully polarized, $\zeta=1$ (bottom) HEG at $r_s=1$, 2 and 40 along with RPIMC data and BDHC fit for $\varepsilon_{\mathrm{xc}}$. }
\label{Fxc-Exc-Hartr-vs-t}
\end{figure*}

Figure \ref{Fxc-Exc-Hartr-vs-t} 
shows XC free, $f_{\rm xc}$, and internal, $\varepsilon_{\rm xc}$, energies 
per particle
from Fit A  for $r_{\mathrm{s}}=1$, 2 and 40
over  $0.01 \le t \le 1000$, with the  $\varepsilon_{\mathrm{xc}}$ energies per
particle compared to the RPIMC data. 
The $\varepsilon_{\mathrm{xc}}$ calculated with the BDHC form
\cite{Brown.fit} also is shown. 
The results for Fit A agree as well with the RPIMC data  
as the BDHC fit.  The $t \rightarrow 0$ limit of the entropic contribution
of course is zero, so the fits and the RPIMC data converge to the $t = 0$ 
$\varepsilon_{\mathrm{xc}}$ value. The zero-$T$ 
unpolarized equilibrium density, $r_{\mathrm s} = 4.19$, from our fit is 
identical with the
value obtained by Perdew and Wang \cite{PW92}. The high-$T$ limit is determined 
by Eq.\ (\ref{highTlim}) for all the $f_{\mathrm {xc}}$ functionals. 
Note that we do not attempt to have our fit describe ordered phases 
({\it e.g.} Wigner crystal) at large $r_{\mathrm s}$.  To do so would
be an unwarranted extrapolation of the RPIMC data.       
Additional comparisons are in the Supplemental Material \cite{SM}.

We now turn to intermediate polarizations.  
In principle the XC functional has separate exchange and correlation 
contributions. At $T = 0$ K, exact 
spin scaling \cite{Oliver.Perdew.1979} defines the X 
functional for arbitrary polarization in terms of the unpolarized one. 
The argument can 
be extended straightforwardly to $T > 0$ K; see the Supplemental Material \cite{SM}.  
No corresponding  
exact result is known for interpolating the C contribution between
$\zeta = 0$ and $\zeta =1$, so approximate forms are used. 
Moreover, it is convenient computationally to use an XC functional
rather than separate X and C contributions.  We used such
a form,
%
%
%
\begin{align}
f_{\mathrm{xc}}(r_{\mathrm{s}},T,\zeta)
&= f_{\mathrm{xc}}^{0}(r_{\mathrm{s}},t) \nonumber \\
+&%
\Big[f_{\mathrm{xc}}^{1}(r_{\mathrm{s}},2^{-2/3}t)-f_{\mathrm{xc}}^{0}(r_{\mathrm{s}},t)\Big]%
\phi(r_{\mathrm{s}},t,\zeta) 
\; ,
\label{fxc-interp}
\end{align}
with 
$\phi(r_{\mathrm{s}},t,\zeta)$ the polarization interpolation
function and $t$ on the right hand side chosen systematically to be
that of the unpolarized case, $t=T/T_F^{\zeta=0}$, as well as in 
Eqs.\ (\ref{phiPDW}) - (\ref{fx-zeta2}) below.  
At $T=0$ K\cite{DreizlerGrossBook}
\be
\phi(\zeta) = \frac{(1+\zeta)^{\alpha} %
+(1-\zeta)^{\alpha} -2}
{2^{\alpha}-2} \, ,
\label{phiTeqZero}
\ee
with $\alpha = 4/3$.  Perrot and Dharma-wardana 
\cite{PDW2000} developed a 
finite-$T$ generalization, $\phi(r_{\mathrm{s}},t,\zeta)$, by replacing 
the exponent $\alpha = 4/3$ with a function, $\alpha (r_{\mathrm{s}},t)$, 
as follows:
\bea
\alpha(r_{\mathrm{s}},t)&=&2-g(r_{\mathrm{s}})\exp\{-t\lambda(r_{\mathrm{s}},t)\} %
\nonumber\\
g(r_{\mathrm{s}})&=&\frac{g_1+g_2r_{\mathrm{s}}}{1+g_3r_{\mathrm{s}}} %
\nonumber\\
\lambda(r_{\mathrm{s}},t)&=&\lambda_1+\lambda_2tr_{\mathrm{s}}^{1/2} \;.
\label{phiPDW}
\eea
Their parametrization used classical map hypernetted chain data for the
HEG and proper behavior as $T \rightarrow 0$ K.   We have
reparametrized $\phi(r_{\mathrm{s}},t,\zeta)$ using the more recent 
$T = 0$ K QMC data (which includes 
intermediate polarizations $\zeta=0.34$, 0.66) \cite{SND.2013} along with the 
CHNC data for intermediate $\zeta=0.6$ in Table IV of Ref.\ \onlinecite{PDW2000}.
(Observe that this is the only use of those CHNC data in this work.)  
The result is a modest improvement for $T=0$ K. The new  
parameter values are in Table \ref{table:intparam}. 
The value of $g_1$ is fixed from the condition that 
$\lim_{r_{\mathrm{s}}\rightarrow 0} \phi(r_{\mathrm{s}},t=0,\zeta)=\phi(\zeta)$.   
The revised $\phi(r_{\mathrm{s}},t,\zeta)$ depends weakly 
on $t$ for all  $r_{\mathrm{s}}$ and $\zeta$.

\begin{table}[h]
\caption{Parameters for the polarization interpolation function given in Eqs.\ 
(\ref{phiTeqZero}) and (\ref{phiPDW}).}
\label{table:intparam}
\begin{ruledtabular}
\begin{tabular}{llll}
   & $\nu=1$ & $\nu=2$ & $\nu=3$ \\
  \hline \\
  $g_\nu$ & 2/3 & -0.0139261 & 0.183208 \\
  $\lambda_\nu$ & 1.064009 & 0.572565 & - \\
\end{tabular}
\end{ruledtabular}
\end{table}

Exact spin interpolation for finite-$T$ exchange 
yields the exchange free energy
\be
f_{\mathrm{x}}(r_{\mathrm{s}},T,\zeta)=\half \Big[
(1+\zeta)^{4/3}f_{\mathrm{x}}^0(r_{\mathrm{s}},t_{\uparrow})+
(1-\zeta)^{4/3}f_{\mathrm{x}}^0(r_{\mathrm{s}},t_{\downarrow})
\Big]
\,,
\label{fx-zeta2}
\ee
where 
$t_{\uparrow/\downarrow}\equiv t(2n_{\uparrow/\downarrow},T)=
2k_BT/[3\pi^2(2n_{\uparrow/\downarrow})]^{2/3}$,
and $n_{\uparrow/\downarrow}=(1\pm\zeta)n/2$. 
(Note that $f_{\mathrm{x}}^\zeta$ for
$\zeta = 0,1$ given by Eq.\ (\ref{fx}) also is given both analytically 
as a Fermi integral and tabulated in Ref.\
\onlinecite{Brown.PRL} Supplementary Material as $E_{x,HF}$.)
Thus the correlation free energy can be found from 
Eqs.\ (\ref{fxc-interp}) and (\ref{fx-zeta2}) to be 

\be
f_{\mathrm{c}}(r_{\mathrm{s}},T,\zeta) =
f_{\mathrm{xc}}(r_{\mathrm{s}},T,\zeta)-f_{\mathrm{x}}(r_{\mathrm{s}},T,\zeta)
\,.
\label{fc}
\ee
To test the $T\rightarrow 0$ K limit of our interpolation, we calculated 
the correlation energy per particle 
\be
\varepsilon_{\mathrm{c}}(r_{\mathrm{s}},\zeta) \equiv f_{\mathrm{c}}(r_{\mathrm{s}},0,\zeta)=
f_{\mathrm{xc}}(r_{\mathrm{s}},0,\zeta)-f_{\mathrm{x}}(r_{\mathrm{s}},0,\zeta)
\,,
\label{e_c}
\ee
where 
$f_{\mathrm{x}}(r_{\mathrm{s}},0,\zeta)\equiv\varepsilon_{\mathrm{x}}(r_{\mathrm{s}},\zeta)$ 
is the LSDA X energy per particle. Comparison with the Perdew-Zunger 
(PZ) LSDA \cite{PZ81} and QMC simulation data shows excellent agreement  
as a function of $\zeta$ for $r_{\mathrm{s}}=$ 0.25, 0.5, 1, 2, 3, 5, 10, and 20,  
with the maximum relative difference between Eq.\ (\ref{e_c}) and the PZ 
correlation energy about 4\% at $r_{\mathrm{s}}=0.25$ and 0.5. 
(Also see  Supplemental Material \cite{SM}.) 

In sum, we have extracted the XC free energy for the finite-$T$ HEG from
the RPIMC data, parametrized it in a form 
with exact asymptotic limits ($r_s\ll 1$, $t = 0$,  and $t\gg1$)
for both the spin unpolarized and fully polarized 
cases, and provided a $T$-dependent interpolation for intermediate
polarizations.  The result, Eqs.\ (\ref{fit2})-(\ref{e}) and 
(\ref{fxc-interp})-(\ref{phiPDW}) and associated parameters, is 
a proper finite-$T$ extension of the widely used ground-state LSDA.  

{\it Acknowledgments}:
We thank Ethan Brown for helpful correspondence and for providing the 
erratum to Ref.\ \onlinecite{Brown.fit} prior to publication and Paul
Grabowski and Aurora Pribram-Jones for a useful remark.  
We thank the University of Florida Research Computing Group for 
computational resources and technical support. VVK, JD, and SBT were supported 
by U.S.\ Dept.\ of Energy grant DE-SC0002139. TS was supported by 
the Dept.\ of Energy Office of Fusion Energy Sciences (FES).


\end{document}